\documentclass[journal]{IEEEtran}
\usepackage{cite}
\ifCLASSINFOpdf
   \usepackage[pdftex]{graphicx}
\else
   \usepackage[dvips]{graphicx}
\fi
\usepackage[cmex10]{amsmath}
\usepackage[normalem]{ulem}

\renewcommand{\vec}[1]{\boldsymbol{#1}}

\begin{document}
\title{Basic aspects of high-power semiconductor laser simulation}
\author{Hans~Wenzel
\thanks{H. Wenzel is with the Ferdinand-Braun-Institut, Leibniz-Institut f\"ur H\"ochstfrequenztechnik, 
Gustav-Kirchhoff-Str. 4, 12489 Berlin, Germany,
e-mail: hans.wenzel@fbh-berlin.de}
\thanks{Manuscript received November 27, 2012; revised January 22, 2013.}}
\markboth{\MakeLowercase{accepted by} Journal Selected Topics Quantum Electronics,~Vol.~19, No.~5, September/October~2013}%
{Wenzel: High-power semiconductor laser simulation}

\IEEEpubid{0000--0000/00\$00.00~\copyright~2013 IEEE}

\IEEEspecialpapernotice{(Invited Paper)}

\maketitle

\begin{abstract}
The aim of this paper is to review some of the models and solution techniques used in the simulation of high-power semiconductor lasers and to address open questions.
We discuss some of the peculiarities in the description of the optical field of wide-aperture lasers.
As an example, the role of the substrate as a competing waveguide in GaAs-based lasers is studied.
The governing equations for the investigation of modal instabilities and filamentation effects are presented
and the impact of the thermal-lensing effect on the spatiotemporal behavior of the optical field is demonstrated.
We reveal the factors that limit the output power  at very high injecton currents based on a numerical solution of the thermodynamic based drift-diffusion equations and elucidate the role of longitudinal spatial holeburning.
\end{abstract}

\IEEEpeerreviewmaketitle

\section{Introduction}

\IEEEPARstart{T}{remendous}
advancements have been achieved in the development
of high-power semiconductor lasers during the last two decades,
due to improved cavity design, crystal growth, facet passivation, and device cooling technologies 
\cite{welch, diehl, behringer2007, lim, landolt, crump2012SemiSemi}.
The output power has increased by one order of magnitude, 
the spectral linewidth has decreased thanks to Bragg gratings integrated into the cavity, 
and the beam quality has greatly improved thanks to tapered gain regions.
Semiconductor lasers are under the way not longer to serve only as optical pumps for solid-state lasers
but to replace them because of their ease-of-use, compactness, efficiency and high reliability.
However, many of the parameters of high-power semiconductor lasers are still far from what is achievable from solid-state lasers.

For example, the reliable maximum output power is currently limited to values below $20$~W 
from a $100$-$\mu$m stripe width device and a further increase remains an open challenge.
High-power semiconductor lasers are also plagued by 
the appearance of multiple peaks in the lateral far- and near field profiles as well as in the optical spectrum in particular at high injection currents.
These effects are caused by  
modal instabilities or by lasing filaments formed by a self-focusing mechanism where the refractive index locally increases in regions of high optical intensity.

In order to assess the root causes of these limitations physics-based modeling and numerical simulation
is more important than ever.
The aim of this paper is to review some of the models and solution techniques used in the simulation of high-power semiconductor lasers and to address open questions.

The paper is organized as follows.
We will start in Section II with a model for the optical field and introduce concepts such as the paraxial parabolic equation, roundtrip operator and modes. 
In Section III we discuss the modeling of the nonlinear interaction between the optical field and the injected carriers in high-power lasers leading to a non-stationary behavior of the field pattern.
The simulation of the stationary electro-optical characteristics of high-power lasers using the thermodynamic based drift-diffusion (or energy transport) model will be presented in Section IV.
We end with a summary and outlook. 

\IEEEpubidadjcol

\section{Description of the optical field}
\label{sect2}

\subsection{Paraxial parabolic equation}

The optical field in edge-emitting semiconductor lasers can be well described by the paraxial parabolic equations \cite{siegman1986}
\begin{multline}
\label{DTWE}
-\frac{ik_0nn_{\text{g}}}{c\beta_0}\frac{\partial E^{\pm}}{\partial t}
\mp i\frac{\partial E^{\pm}}{\partial z}
+ \frac{1}{2 \beta_0}\mathbf{\Delta}_t E^{\pm} \\
+ \frac{k_0^2n^2-\beta_0^2}{2\beta_0} E^{\pm} 
+ \frac{k_0^2\xi^{\pm}}{2 \beta_0}E^{\mp}
= 0
\end{multline}
for the right and left traveling waves $E^+$ and $E^-$, respectively. 
Eq. (\ref{DTWE}) can be derived from Maxwell equations with the Ansatz 
\begin{multline}
\label{ansatz}
\vec{E}(\vec{r}_t,z,t)=\vec{e}_x\big[E^+(\vec{r}_t,z,t)\text{e}^{-i\beta_0z}\\ +E^-(\vec{r}_t,z,t)\text{e}^{+i\beta_0z}\big]\text{e}^{i\omega_0t}
\end{multline}
for the transverse electric field in order to remove the rapid variations with respect
to the time $t$ and the coordinate $z$ along the cavity axis, and a corresponding Ansatz for the complex dielectric function, 
\begin{multline}
\label{dielectric}
\varepsilon(\vec{r}_t,z,t,\omega)|_{\omega_0}= n^2(\vec{r}_t,z,t)\\
+\xi^+(\vec{r}_t)\text{e}^{-i2\beta_0z}
+\xi^-(\vec{r}_t)\text{e}^{+i2\beta_0z}.
\end{multline}
In (\ref{DTWE}-\ref{dielectric}), $\vec{r}_t=(x,y)$ are the transverse coordinates, $\mathbf{\Delta}_t$ is the transverse Laplacian, $\omega_0$ and $\beta_0$ are the real-valued reference frequency and reference propagation factor, respectively, $k_0=\omega_0/c=2\pi/\lambda_0$ with $\lambda_0$ being the reference vacuum wavelength and $c$ the vacuum velocity of light, $n_{\text{g}}=\partial(\omega n)/\partial \omega|_{\omega_0}$ is the group index and $\xi^{\pm}$ are the Fourier coefficients of a Bragg grating integrated into the cavity.
We have further assumed a TE-polarized field with $\vec{e}_x$ being the unity vector and that
the dielectric function varies slowly with respect to $x$. 
The distribution of the complex-valued refractive index $n$ (which may include a time dependence much slower than the variation given by $1/\omega_0$) can be written as
\begin{equation}
n=n_{\text{r}}+i\frac{g-\alpha}{2k_0}
\end{equation}
with $g$ being the gain due to inter band transition and $\alpha$ the absorption coefficient due to intraband transitions (\textit{e.g.} free carrier absorption).

The temporal dispersion of the dielectric function has been taken into account only up to the first order in the real part of the refractive index $n$, higher order dispersion and the dispersion of the imaginary part have to be properly added (see Section \ref{sect3}).
We should note that in a numerical evaluation of (\ref{DTWE}) the factor in front of the time-derivative must be approximated by 
\begin{equation}
\label{groupindex}
\frac{k_0nn_{\text{g}}}{c\beta_0} \approx \frac{1}{v_\text{g}}
\end{equation} 
with a real-valued group velocity $v_\text{g}$.

Eqs. (\ref{DTWE}) must be supplemented by appropriate boundary conditions.
At the plane facets of the laser
\begin{equation}
\begin{gathered}
\label{refl}
E^+(0) - r_0E^-(0)=0,\\
E^-(L) - r_Le^{-i2\beta_0L}E^+(L)=0
\end{gathered}
\end{equation}
hold at $z=0$ and $z=L$, respectively.
We should mention that in the paraxial approximation the facet reflectivities $r_0$ and $r_L$ are input parameters which have to be calculated in advance.

At the transverse boundary denoted by $\Gamma$ one can assume, for example, decaying fields or a perfect electric wall,
\begin{equation}
\label{TBC}
\lim_{|\vec{r}_t|\to\infty}E^{\pm}=0\quad\text{or}\quad
E^{\pm}|_{\vec{r}_t \in \Gamma}=0,
\end{equation}
respectively.
If only a part of the cross-section of the cavity is simulated, a non-reflecting or transparent boundary condition has
to be used which models the fact that only outgoing waves should be present.
This boundary condition can be formulated only in operator form for the general case,
\begin{equation}
\label{outgoing}
\frac{\partial E^{\pm}}{d\vec{n}}|_{\vec{r}_t \in \Gamma} =-i\mathbf{D}E^{\pm}|_{\vec{r}_t \in \Gamma}
\end{equation}
where $\mathbf{D}$ is the operator for a so-called Dirichlet-to-Neumann map \cite{antoine}.
We will consider below in Subsection \ref{subsectE} an example for a one-dimensional case.
For spatially and temporally constant $n$, $\mathbf{D}$ can be obtained by a factorization of (\ref{DTWE})\cite{moore1988}.
A very popular method to implement a boundary condition of type (\ref{outgoing}) is the introduction of a so-called perfectly matched layer (PML) \cite{oskooi}.
However, in the frequency domain it results in a large number of spurious modes \cite{tischler}.

\subsection{Cavity modes}

The cavity modes are 
time-periodic solutions of the form $E_m^{\pm}\text{exp}(i\Omega_m t)$ and obey the equations
\begin{multline}
\label{cavitymodes}
\frac{k_0nn_{\text{g}}\Omega_m}{c\beta_0}E_m^{\pm} 
\mp i\frac{\partial E_m^{\pm}}{\partial z}
+ \frac{1}{2 \beta_0}\mathbf{\Delta}_t E_m^{\pm} \\
+ \frac{k_0^2n^2-\beta_0^2}{2\beta_0}E_m^{\pm} 
+ \frac{k_0^2\xi^{\pm}}{2 \beta_0}E_m^{\mp}
= 0
\end{multline}
subject to the boundary conditions (\ref{refl})-(\ref{outgoing}).
The non-trivial solutions of (\ref{cavitymodes}) may dependent on time via the time-dependence of $n$.
The complex-valued relative mode frequencies $\Omega_m$ are the eigenvalues and the mode profiles $E_m^{\pm}(\vec{r}_t,z,t)$ are the eigenfunctions of (\ref{cavitymodes}).
Some mathematical properties and their physical impact will be discussed below.
The real parts of $\Omega_m$ give the wavelengths relative to the reference wavelength $\lambda_0$,
\begin{equation}
\label{deltalambda}
\Delta \lambda_{m}=\frac{d\lambda}{d\omega}\Big|_{\lambda_0}\text{Re}(\Omega_{m})
\end{equation}
and the imaginary parts describe the damping of the modes.
For a passive cavity, $\text{Im}(\Omega_m)>0$ must hold.
Lasing modes of an active cavity are distinguished by vanishing damping, $\text{Im}(\Omega_m)=0$, due to the balance of the outcoupling and internal losses and the gain.

The cavity modes fulfill the orthogonality relation
\begin{equation}
\label{Orthogonality}
\int nn_\text{g}\left[E_m^+E_{m'}^-+E_m^-E_{{m'}}^+\right]dxdydz=0\quad\text{for}\quad m\neq {m'}
\end{equation}
which is proven in the Appendix.
It is similar to corresponding orthogonality relations for cavity modes derived in \cite{wenzel1994} and \cite{batrak} from Maxwell and Helmholtz equations, respectively.

Note that the integral (\ref{Orthogonality}) does not define a scalar product \cite{wenzel1996jqe}.
In fact, (\ref{cavitymodes}) defines a non-Hermitian eigenvalue problem, because the frequencies $\Omega_m$ are complex-valued.
It is well-known from the theory of non-Hermitian operators, that in dependence of some parameter(s) so-called \lq\lq exceptional points\rq\rq\ exist where the eigenvalues (both real and imaginary parts) cross and the eigenfunctions become identical \cite{heiss}.
Mode degeneracies related to exceptional points have been discovered for unstable laser cavities,
\textit{cf.} \cite{berry} and the references therein, multi-section lasers \cite{wuensche1995, liertzer} and complex planar waveguides \cite{ctyroky}.
Due to the mode degeneracy, at an exceptional point the system of eigenfunctions is no longer complete, but a system including the generalized eigenfunctions is, \textit{cf.} \cite{rehberg}. 

At the exceptional point, the modes remain orthogonal in the sense of (\ref{Orthogonality}), so that the integral
$\int nn_\text{g}E_m^+E_m^-dxdydz$\ 
vanishes.
As a consequence, Petermann's $K$ factor \cite{petermann}
\begin{equation}
\label{K}
K_m=\left|\frac{\int nn_\text{g}\left[|E_m^+|^2 +|E_m^-|^2\right]dxdydz}
{2\int nn_\text{g} E_m^+E_m^-dxdydz}\right|^2,
\end{equation}
approaches infinity.
It is known, that the $K$ factor causes an enhancement of the spontaneous emission, which results in a broadening of the spectrum of a multi-longitudinal-mode laser and the spectral linewidth of a single mode as summarized in \cite{siegman1995}. 
Its time-dependence is needed to explain the dynamical behavior of multi-section lasers \cite{bandelow1996}.

\subsection{Beam propagation method and roundtrip operator}

For a solution of (\ref{cavitymodes}) it is convenient to introduce the operator
\begin{equation}
\mathbf{H}(\vec{r}_t,z)=
\frac{1}{2 \beta_0}\mathbf{\Delta}_t+\Delta\beta
\end{equation}
with the abbreviation
\begin{equation}
\Delta\beta=\frac{k_0^2n^2-\beta_0^2}{2\beta_0}.
\end{equation}
If $\xi^\pm=0$ holds (Fabry-Perot cavity), and if the index $n$ depends only on the transverse coordinates, $n=n(\vec{r}_t)$, the solution of (\ref{cavitymodes}) can be formally written as
\begin{equation}
\label{BPM}
E_m^\pm(\vec{r}_t,z^\prime)=\text{e}^{\mp i(\frac{\Omega_m}{v_{\text{g}}}+\mathbf{H})(z^\prime-z)} E_m^\pm(\vec{r}_t,z)\quad\text{(FP cavity)},
\end{equation}
using the approximation (\ref{groupindex}).
The numerical evaluation of (\ref{BPM}) is the basis what is known as the beam propagation method (BPM) \cite{maerz, scarmozzino, benson}.
It has been used for the simulation of both passive \cite{sujecki2003} and active \cite{borruel2004, lim, odriozola2009} high-power laser cavities.

For the case of a spatially and temporal constant index, $n=\text{const.}$, (\ref{BPM}) can be evaluated exactly to yield
\begin{multline}
\label{Propagator}
E_m^\pm(\vec{r}^\prime_t,z^\prime)=
\text{e}^{\mp i(\frac{\Omega_m}{v_{\text{g}}}+\Delta\beta)(z^\prime-z)}\\
\times\int G^\pm(\vec{r}^\prime_t-\vec{r}_t,z^\prime-z)E_m^\pm(\vec{r}_t,z)dxdy\quad\text{(FP cavity)}
\end{multline}
with 
\begin{multline}
\label{Green}
G^\pm(\vec{r}^\prime_t-\vec{r}_t,z^\prime-z)=\\
\Theta\big(\pm(z^\prime-z)\big)\left[\sqrt{\pm\frac{i\beta_0}{2\pi(z^\prime-z)}}\right]^2\text{e}^{\mp\frac{i\beta_0|\vec{r}^\prime_t-\vec{r}_t|^2}{2(z^\prime-z)}}
\end{multline}
where $\Theta$ denotes the Heaviside step function.
The integral equation (\ref{Propagator}) together with the propagator (\ref{Green}) is known as Huygen's integral in the Fresnel approximation.
It has been successfully applied for the simulation of laser cavities as outlined in \cite{siegman2000a, siegman2000b}, 
including those of semiconductor lasers  \cite{fukushima, spreemann2011, spreemann2012}. 

Based on (\ref{BPM}) and the boundary conditions (\ref{refl}) it is possible to construct round trip operators $\mathbf{M}^{\pm}$.
One starts at some position $z=z_0$ within the cavity and performs a full roundtrip.
Depending whether we start into forward ($+$) or backward ($-$) directions, the eigenvalue problems
\begin{equation}
\label{roundtrip}
\mathbf{M}^{\pm}E_m^\pm(\vec{r}_t,z_0)=\gamma_m E_m^\pm(\vec{r}_t,z_0)
\end{equation}
are obtained with
\begin{equation}
\gamma_m=\text{e}^{i\frac{2\Omega_m}{v_{\text{g}}}L}
\end{equation}
being the eigenvalues.
The eigenfunctions of $\mathbf{M}^{\pm}$ are the mode distributions $E_m^\pm(\vec{r}_t,z_0)$ at the position $z_0$.

A very popular method for solving (\ref{roundtrip}) is based on the Fox-Li approach, \textit{cf.} \cite{siegman2000a} and the references therein.
The idea is to choose a normalized, more or less arbitrary start distribution $E_m^\pm(\vec{r}_t,z_0)$ 
and to apply the round trip operator $\vec{M}^{\pm}$ recurrently until one arrives (hopefully) at a steady state, 
\textit{i.e.} a distribution that does not change its pattern in a round-trip,
except for a reduction in amplitude and a phase shift which yields $\gamma_m$.
Thereby typically the phase factor $\text{exp}\big(-2i\text{Re}(\Delta\beta)L-2i\beta_0L)\big)$ is omitted.

The Fox-Li approach delivers the mode with the largest $|\gamma_m|$ if converged which is, however, not guaranteed.
From a mathematical point of view, 
it is related to the power (also called vector or von-Mises) iteration to determine the largest eigenvalue of a matrix and the corresponding eigenvector.
It is known, that the generated sequences converge only if the eigenvalue is simple and well separated from the other eigenvalues.
Hence, the algorithm fails if there are cavity modes having identical or nearly identical dampings $\text{Im}(\Omega_m)$. 
Nowadays, there are more stable and efficient methods to solve (\ref{roundtrip}) for passive laser cavities based on Arnoldi or Lanczos iteration.
Recently, an alternative approach based on a finite-element method for solving (\ref{cavitymodes}) directly, without recourse to (\ref{roundtrip}), has been proposed  \cite{altmann}.

Simulations of active laser cavities based on BPMs still utilize the Fox-Li approach \cite{lim, napartovich},
which work quite well for single-transverse mode lasers.
The approach fails for multi-transverse mode lasers and at high output power if nonlinearities such as self-focusing and filamentation start to dominate, \textit{cf.} \cite{sujecki2007} for a recent discussion of the stability of the Fox-Li approach.
The association of the nonconvergence of the Fox-Li iteration with an dynamically unstable laser behavior (as in
\cite{marciante1996}) 
should be done with care.
Instead, for multi-mode high-power lasers, a time-dependent approach based on (\ref{DTWE}) should be preferred, although numerically more challenging.

\subsection{Beyond paraxial approximation}

With improving computer capabilities, direct numerical solutions of Maxwell or Helmholtz equations without recourse to the paraxial approximation become more and more feasible.
For example, in \cite{kalosha} a scalar Helmholtz equation has been solved in the $(x,z)$ (lateral, longitudinal) plane for a high-power laser, restricted, however, to short cavities.
An numerical model based on the Helmholtz operator and a first-order time derivative obtained from Maxwell equations by separating the rapidly varying term $\text{exp}(i\omega t)$ has been used in \cite{berneker} for the simulation of a monolithically integrated master-oscillator power-amplfier.
Another possibility is the expansion in terms of waveguide modes \cite{hocke, bienstman2001} considered in the next Subsection.

\subsection{Waveguide modes}

We can expand the left and right traveling waves in terms of waveguide modes,
\begin{equation}
\label{expansion}
E^{\pm}(\vec{r}_t,z,t)=\sum_\nu A_\nu^{\pm}(z,t)\chi_\nu(\vec{r}_t,z,t),
\end{equation}
which are solution of
\begin{equation}
\label{waveguide}
\mathbf{\Delta}_t\chi_\nu
+ \left[k_0^2n^2-\beta_\nu^2\right]\chi_\nu=0
\end{equation}
subject to the boundary conditions (\ref{TBC}). 
The waveguide modes are exact solutions of Maxwell equations of the form
$\vec{E}_\nu(\vec{r}_t)\text{exp}(i\omega t-i\beta_\nu z)$ if the dielectric function is translationally invariant along $z$ \cite{pomplun}.

Eq. (\ref{waveguide}) is in general a non-hermitian eigenvalue problem for the complex propagation factor $\beta_\nu$.
An overview of several methods to solve (\ref{waveguide}) can be found in \cite{scarmozzino}.
Expansion into waveguide modes allows an exact treatment of waveguide discontinuities such as
Bragg gratings \cite{fricke} 
or laser facets \cite{derudder} in order to calculate their reflectivities.
In what follows we introduce an approximate solution of (\ref{waveguide}).

\subsection{Effective index method}
\label{subsectE}

For a nearly planar waveguide, \textit{i.e.} if the mode profile $\chi(x,y)$ does not vary
strongly along the $x$ direction, (\ref{waveguide}) can be solved perturbationally \cite{kato}.
We choose
\begin{equation}
\label{EIM}
\chi(x,y)=\Phi(x,y)\phi(x)
\end{equation}
where the dependence of $\Phi(x,y)$ on $x$ is only parametrically.
If $\Phi(x,y)$ is the solution of the eigenvalue problem
\begin{equation}
\label{vertical}
\frac{d^2\Phi}{dy^2}+k_0^2\left[\bar{n}^2-n_{\text{eff}}^2\right]\Phi=0
\end{equation}
with 
\begin{equation}
\int_{-\infty}^{\infty} \Phi^2dy = 1,
\end{equation}
then $\phi(x)$ obeys the eigenvalue problem
\begin{equation}
\label{lateral}
\frac{d^2\phi}{dx^2}+\left[k_0^2\left(n_{\text{eff}}^2+\Delta n_{\text{eff}}^2\right)-\beta^2\right]\phi=0
\end{equation}
with
\begin{equation}
\label{neff}
\Delta n_{\text{eff}}^2=\int\left[n^2-\bar{n}^2\right]\Phi^2dy.	
\end{equation}
In (\ref{vertical}), $\bar{n}(x,y)$ is a typically real-valued chosen refractive-index distribution of a reference waveguide and $n_{\text{eff}}(x)$ is the so-called effective index at position $x$.
The correction $\Delta n_{\text{eff}}$ given by (\ref{neff}) could include \textit{e.g.} modifications of the index due to carrier density and temperature effects as well as the imaginary part.
The approach sketched above is called \lq\lq effective-index method\rq\rq\ in the semiconductor laser community.
Eq. (\ref{vertical}) is widely used in the design of high-power lasers, because it allows the optimization of the layer structure with respect to optical confinement, far-field divergence and optical losses.
In what follows we will consider an example.

\subsection{Worked example}
\label{subsectF}

\begin{figure}[!t]
\centering
\includegraphics[width=2.5in]{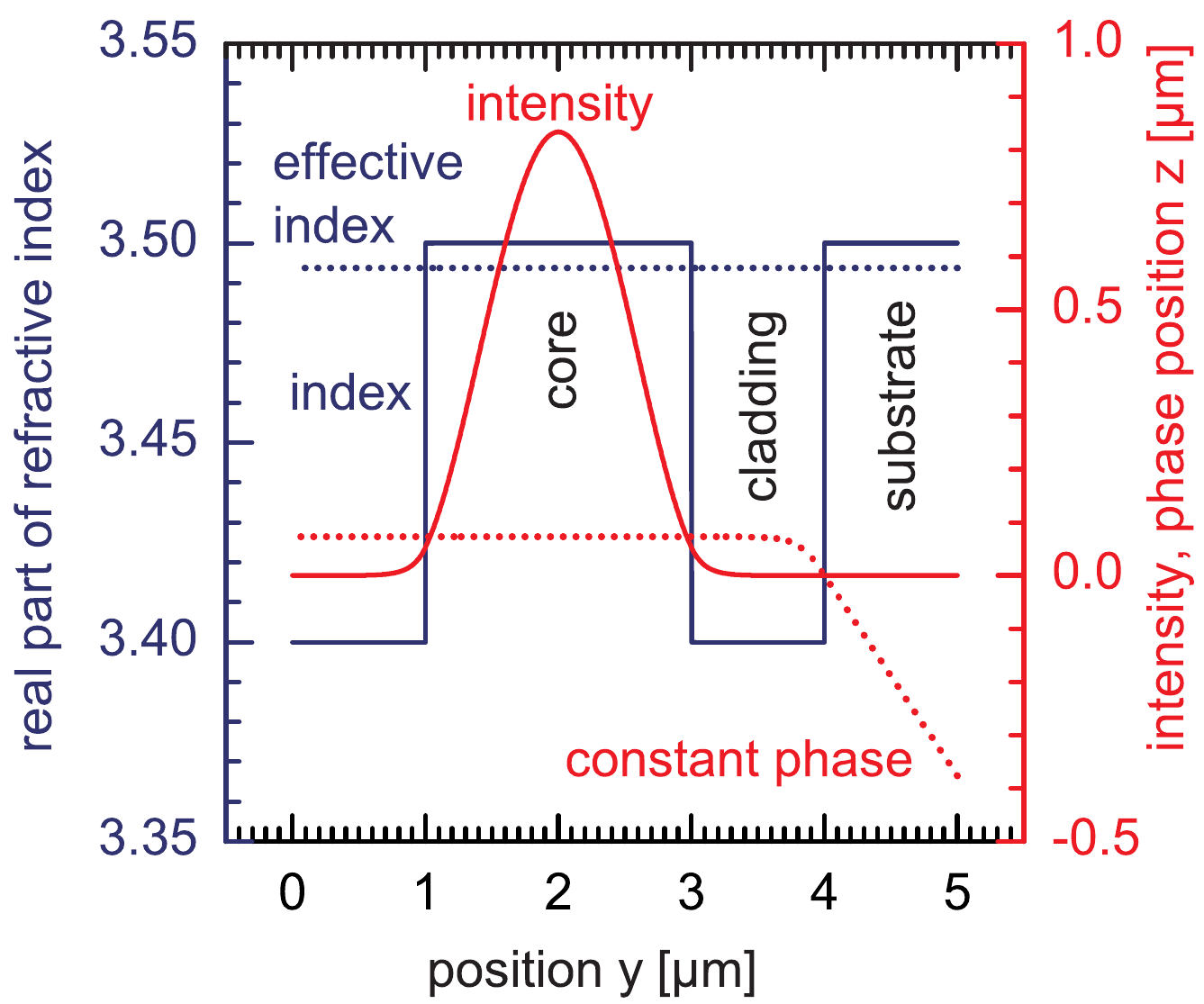}
\caption{Profiles of real part of refractive index(blue solid, left axis), intensity (red solid, right axis) and constant phase (red dotted, right axis) of the leaky waveguide under study. 
The real part of the effective index is also shown (blue dotted, left axis.)}
\label{vnf}
\end{figure}

Lasers grown on GaAs substrates, suffer from a leakage of the lasing mode into the substrate, or a coupling of the lasing mode with substrate modes because the refractive index of the substrate is typically larger than the effective index of the lasing mode.
Although this effect is well-known for a long time \cite{bogatov, eliseev1994, avrutsky}, it has been recently re-discovered for lasers grown on GaN substrates \cite{laino}.
A notably exception are lasers with a broad GaAs waveguide core and a multi-quantum well active region emitting around $1100~$nm \cite{erbert}.

Assuming an infinite thick substrate, the associated losses can be determined by solving (\ref{vertical}) subject to the condition
\begin{equation}
\label{radiation}
\frac{d\Phi}{dy}=-ik_0\sqrt{n_{\text{sub}}^2-n_{\text{eff}}^2}\Phi
\end{equation}
at the boundary between cladding and substrate.
Eq. (\ref{radiation}) is an example of a boundary condition of type (\ref{outgoing}).
Note that the correct branch of the complex square root function has to be chosen in order to ensure an outgoing wave.
As a consequence of (\ref{radiation}), the effective index $n_{\text{eff}}$ becomes complex valued, even though there is no absorption in the waveguide.
The imaginary part of the effective index describes the loss of the modes through radiation into the substrate.
If there is no absorption in the substrate or gain in an active layer, the field diverges with increasing distance from the cladding layer. 
If the absorption or gain is sufficiently large, this so-called leaky mode becomes a proper, albeit complex, guided mode.

Simulation tools using a PML boundary condition often fail to calculate the radiation losses of leaky waveguides correctly \cite{bienstman2006}.
Using a transfer-matrix based method, the radiation loss can be exactly calculated by solving a complex-valued transcendental equation
\cite{wenzel1990, petracek}.
We will give here the results of a worked example which can be used as a benchmark.
Fig. \ref{vnf} shows the profile of the index and the profile of the intensity of the mode with the highest effective index for a vacuum wavelength of $\lambda_0=1~\mu$m.

\begin{figure}[!t]
\centering
\includegraphics[width=2.8in]{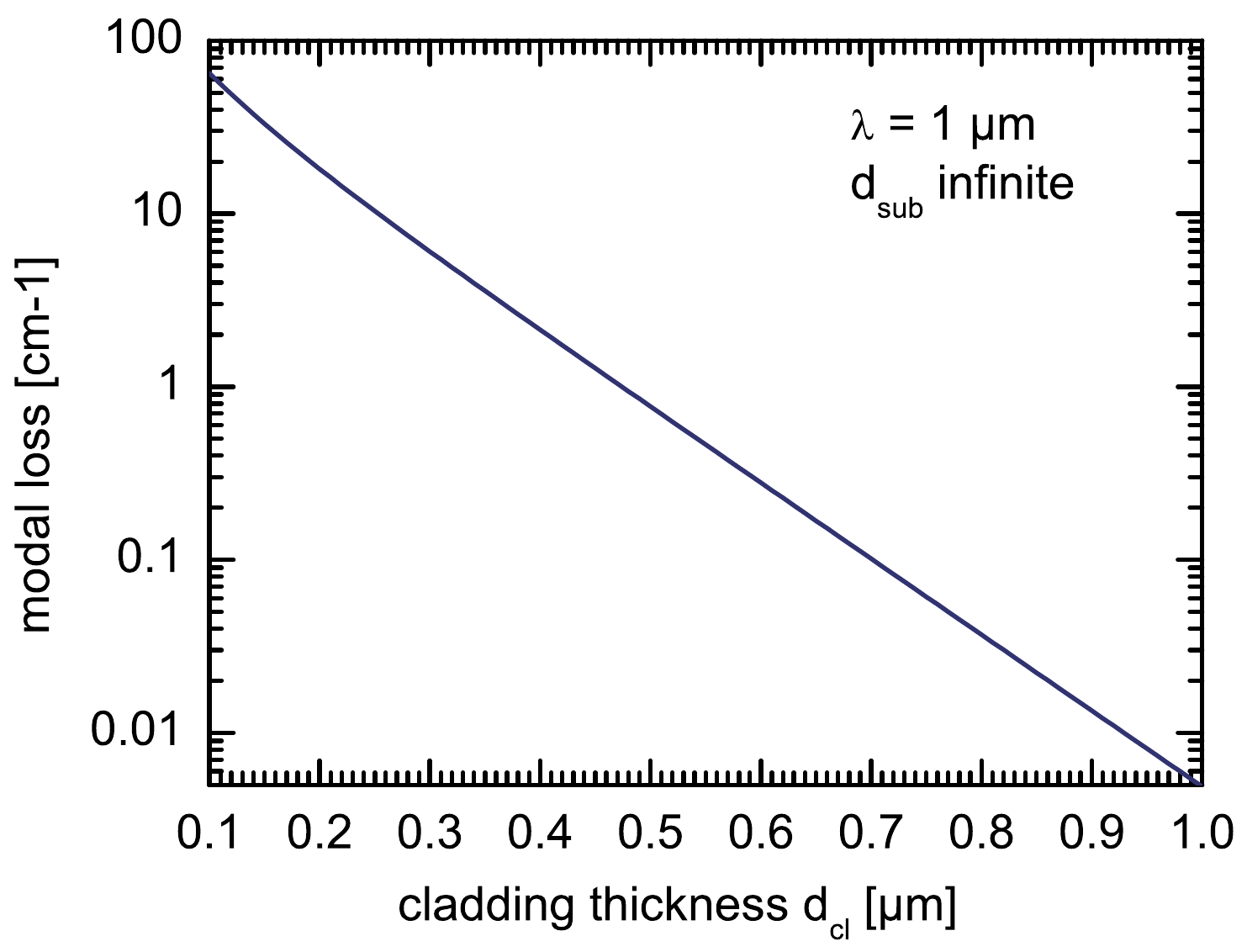}
\caption{Modal losses due to radiation into an infinitely thick substrate versus the thickness of the cladding layer. The vacuum wavelength is $\lambda_0=1~\mu$m.}
\label{vloss}
\end{figure}

Fig. \ref{vloss} shows the radiation losses in dependence of the thickness of the cladding layer between waveguide core and substrate. The dependence is exponential as to be expected, except for small values of $d_{\text{cl}}$, where the mode becomes strongly leaky.
Due to the fact that the radiation loss increases with the effective index, a careful adjustment of the thickness of the cladding layer between substrate and waveguide core can be used to discriminate higher order vertical modes \cite{erbert}.    

\begin{figure}[!t]
\centering
\includegraphics[width=2.8in]{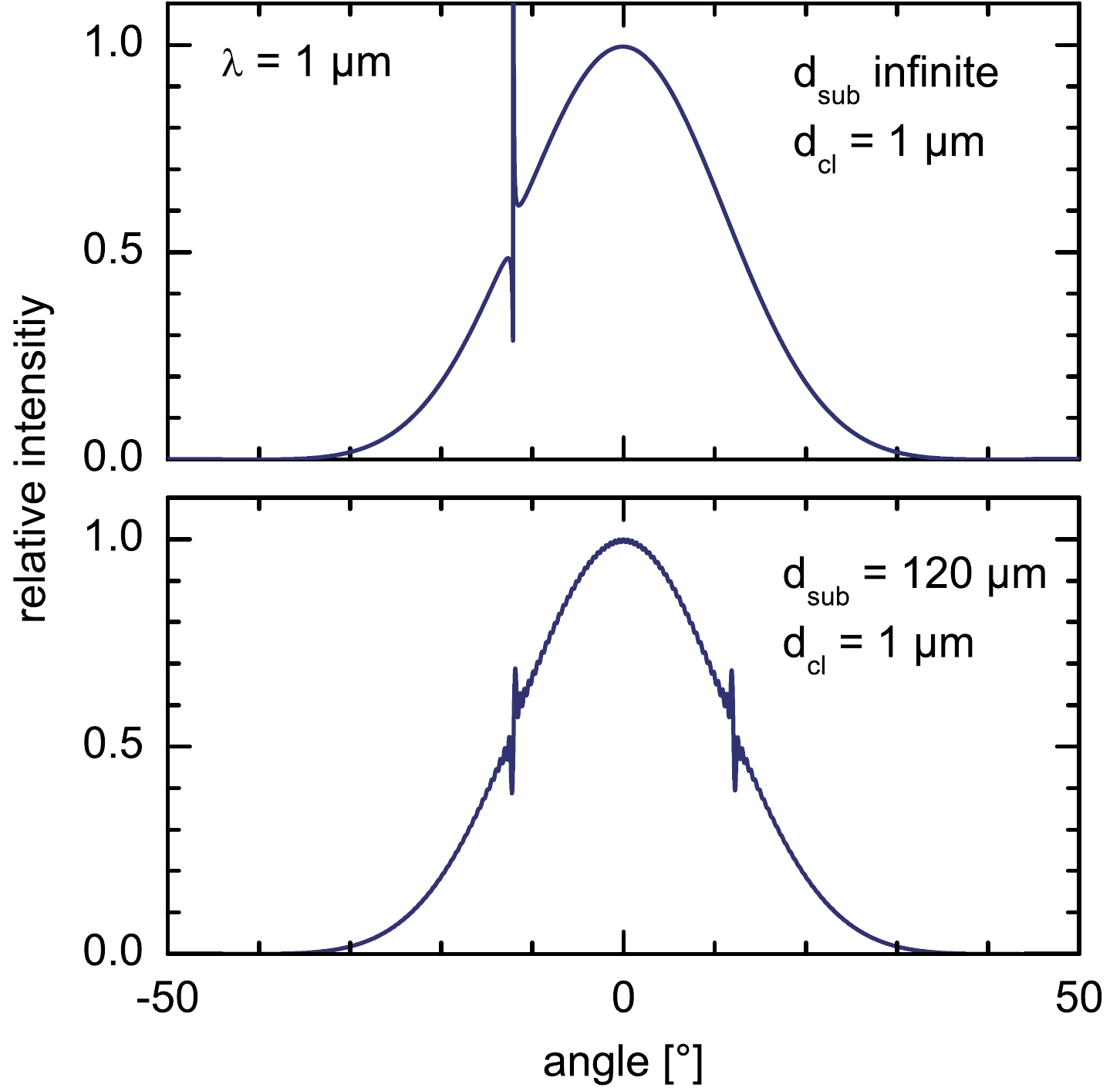}
\caption{Vertical profiles of the far field intensities for a substrate infinitely thick (top) and with a finite thickness (bottom).}
\label{vff}
\end{figure}

A second consequence of the optical leakage is the appearance of additional peak(s) in the profile of the far field intensity
\begin{equation}
\label{FF}
P_{FF}(\Theta)\propto \text{cos}^2(\Theta)\Big|\int\Phi(y)\text{e}^{ik_0\text{sin}(\Theta)y}dy\Big|^2
\end{equation}
in dependence on the vertical divergence angle $\Theta$. 
If the mode profile in the substrate
\begin{equation}
\Phi(y)\propto\text{e}^{-ik_0\sqrt{n_{\text{sub}}^2-n_{\text{eff}}^2}y}
\end{equation}
is inserted into (\ref{FF}), there is a resonance if 
\begin{equation}
\label{resonance}
\text{sin}^2(\Theta_r)=n_{\text{sub}}^2-n_{\text{eff}}^2
\end{equation}
holds.
Depending whether a substrate with an infinite (Fig. \ref{vff}, top) or finite (Fig. \ref{vff}, bottom)  
thickness is assumed, one resonance or two resonances, respectively, will be observed in the far field profile.
The magnitude and width of the peaks depend on the absorption in the substrate and the bottom surface roughness and metalization.
The additional peak(s) in the farfield can be observed if the condition 
$0<n_{\text{sub}}^2-n_{\text{eff}}^2<1$ holds.
If $n_{\text{sub}}^2-n_{\text{eff}}^2>1$, the internal propagation angle $\Theta_i=\text{acos}(n_{\text{eff}}/n_{\text{sub}})$,
which is the angle of the normal of the line of constant phase shown in Fig. \ref{vnf} with respect to the $z$ axis,
is larger than the critical angle of total reflection and no peak appears.

\begin{figure}[!t]
\centering
\includegraphics[width=2.8in]{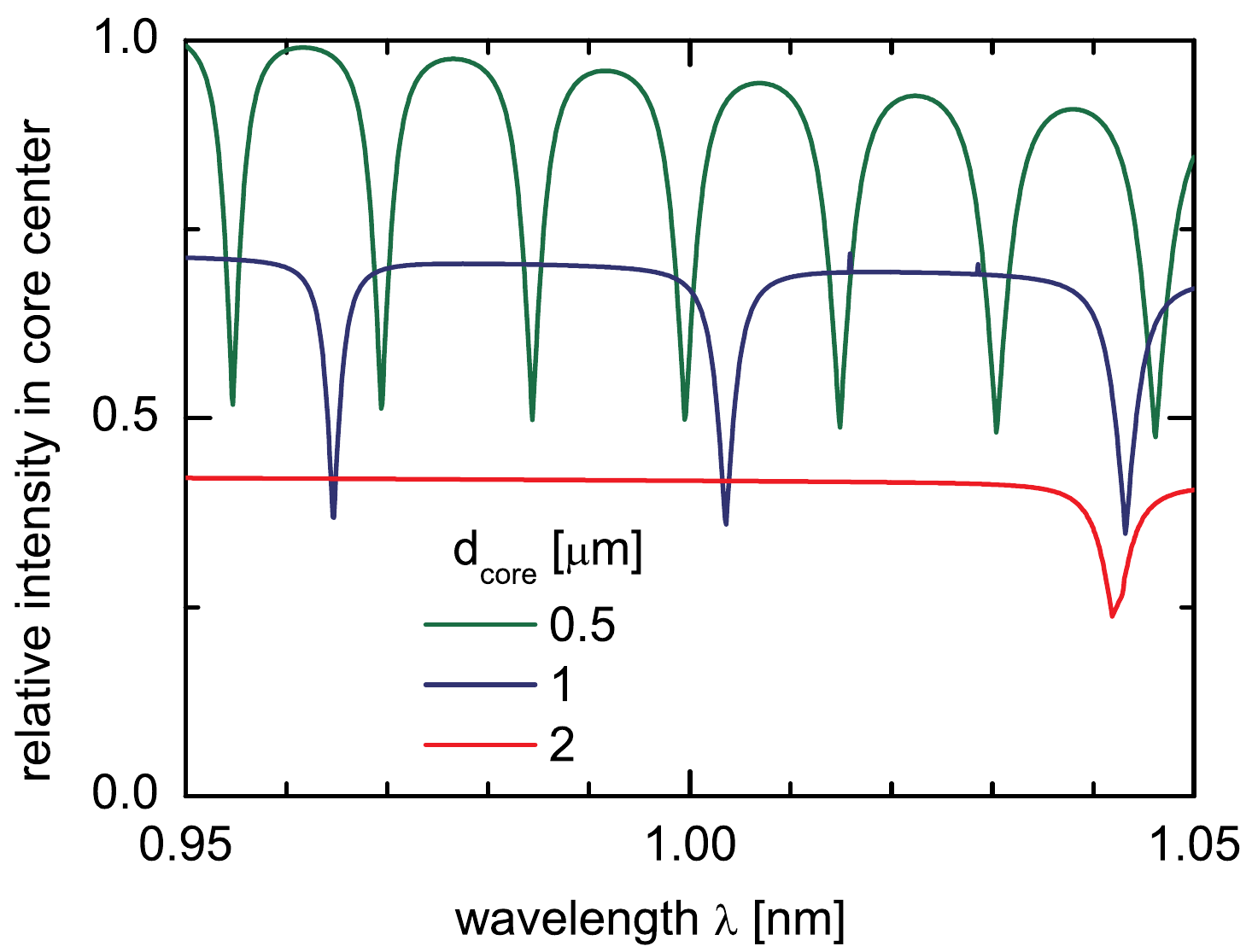}
\caption{Relative intensity in the waveguide core versus vacuum wavelength for different width of the waveguide core as indicated.}
\label{vgam}
\end{figure}

A third consequence of the coupling of the lasing mode with substrate modes is a modulation of the intensity in the center of the waveguide core (or the optical confinement factor) in dependence on the wavelength as shown in Fig. \ref{vgam}.
The modulation period depends strongly on the width of the waveguide core because the smaller the core
the larger the wavelength dependence of the index of the mode confined to the core.
This effect results in a modulation of the modal gain and hence in a corresponding modulation of the spectrum of the amplified spontaneous emission \cite{avrutsky, witzigmann2007} and the optical spectrum above threshold \cite{horie}.

It should be noted that the GaAs p-contact layer causes mode coupling phenomena, too, if the thickness is not properly chosen \cite{eliseev1994}.

\subsection{Longitudinal modes}

If we insert (\ref{expansion}) into (\ref{cavitymodes}) and neglect the coupling of different transverse modes due to a spatially varying group index, temporally varying index or spatially varying index perturbations, 
the mode amplitudes obey
\begin{equation}
\label{TME1}
\frac{i \Omega_{\nu k}}{v_{\text{g},\nu}}A_{\nu k}^{\pm}  \pm \frac{d A_{\nu k}^{\pm}}{dz}
+ i\Delta\beta_\nu A_{\nu k}^{\pm}
+ i\kappa_\nu^{\pm}A_{\nu k}^{\mp}
= 0,
\end{equation}
with the abbreviation
\begin{equation}
\Delta\beta_\nu = \frac{1}{2\beta_0}\left[\beta_\nu^2-\beta_0^2\right].
\end{equation}
For a Fabry-Perot laser, where the coupling coefficients 
\begin{equation}
\kappa_\nu^{\pm}= \frac{k_0^2\int\xi^{\pm}E_\nu^2\;dxdy}{2\beta_0 \int E_\nu^2\;dxdy}
\end{equation}
vanish, Eq. (\ref{TME1}) can be analytically solved subject to the boundary conditions (\ref{refl}).
From the complex mode frequencies and (\ref{deltalambda}) the wavelengths of the modes relative to the reference wavelength can be determined to
\begin{equation}
\label{deltalambdaFP}
\Delta \lambda_{\nu k}^{\text{FP}}=
-\frac{\lambda_0^2}{2L\pi n_{\text{g},\nu}}\Big[ \frac{\varphi_0+\varphi_L}{2} + \pi k - L\beta_0 - L\text{Re}(\Delta\beta_\nu)\Big]
\end{equation}
where $k$ is the longitudinal mode index and $\varphi_0$ and $\varphi_L$ are the phases of the reflectivities.
The spacing between the wavelengths of different transverse modes belonging to the same longitudinal mode is approximately given by
\begin{equation}
\label{ModenAbstand}
\lambda_{\nu k}-\lambda_{\nu'k}
\approx
\frac{\lambda_0}{n_{\text{g},\nu}}\left(n_{\text{mod},\nu}-n_{\text{mod},\nu'}\right)
\end{equation}
with $n_{\text{mod},\nu}=\text{Re}(\beta_\nu)/k_0$.

Several cases can be distinguished depending on the difference of the modal indices governing the wavelength spacing of the transversal modes.
For example, in a ridge waveguide laser with a narrow ridge, the spacing of the wavelengths of the lateral modes is in the order of $1$~nm which is much larger than that of the longitudinal modes.
Hence, every lateral mode is associated with a comb of longitudinal modes which can overlap.
In a broad-area laser with a wide ridge or gain region, on the other hand, the spacing of the wavelengths of the lateral modes is typically much smaller than that of the longitudinal modes, so that in a measured optical spectrum each longitudinal mode splits into several peaks belonging to the different lateral modes.

Assuming a 1D waveguide with width $W$ and constant real refractive index $n_{\text{r}}$ surrounded by perfect electric walls, the modal indices are
\begin{equation}
n_{\text{mod},\nu}=\sqrt{n_{\text{r}}^2-\frac{\lambda_0^2}{4W^2}\nu^2}
\quad\text{with}\quad \nu=1,2,3,\cdots.
\end{equation} 
Thus the wavelength spacing with respect to the fundamental mode $\nu=1$ is given approximately
\begin{equation}
\label{wavelengthspacing}
\lambda_{\nu k}-\lambda_{1k}
\approx \frac{\lambda_0^3}{8n_{\text{r}}n_{\text{g}}W^2}(1-\nu^2),
\end{equation}
which has been used by several authors \cite{stelmakh2006, crump2012} for studies of broad-area lasers.
It should be noted that (\ref{wavelengthspacing}) is an approximation, because it neglects the true profile of the 
index and the penetration of the fields into the exterior region.

\section{Time-dependent active cavity simulation}
\label{sect3}

\subsection{Model equations}

In order to assess the multi-mode behavior, modal instabilities and filamentation effects
of wide-aperture semiconductor lasers a time-dependent approach is the most appropriate one
as outlined in Section II. 
Due to computer limitations, until now only the  
vertical-projected equations \cite{ning1997, lichtner}
\begin{multline}
\label{TWE1}
-\frac{i}{v_{\text{g,eff}}}\frac{\partial E^{\pm}}{\partial t} 
\mp i\frac{\partial  E^{\pm}}{\partial z}
+ \frac{1}{2 \beta_0}\frac{\partial^2 E^{\pm}}{\partial x^2} \\
+ \frac{k_0^2\big[n_{\text{eff}}^2+\Delta n^2_{\text{eff}}\big] -\beta_0^2}{2\beta_0} E^{\pm} 
+ \frac{k_0^2\kappa^{\pm}}{2 \beta_0}E^{\mp}
= F_{\text{spont}}
\end{multline}
with
\begin{equation}
\kappa^{\pm}= \frac{k_0^2}{2 \beta_0} \int \xi^{\pm} \Phi^2 dy
\end{equation}
have been dealt with.
They are obtained by introducing an Ansatz like (\ref{EIM}) into (\ref{DTWE}) and adding
a Langevin source $F_{\text{spont}}$ on the rhs describing spontaneous emission.
Dispersion of the imaginary part of the index (optical gain) must be additionally included to eliminate the high-$k$ instability \cite{jakobsen1992},
which can be done on a microscopic level \cite{ning1997SPIE, gehrig}, 
as an effective polarization \cite{ning1997} modeling a Lorentzian shape of the gain,  
with higher order time derivatives \cite{balsamo}, by means of a digital filter \cite{kolesik},
or by a convolution integral
\cite{javaloyes2010}.

Eq. (\ref{TWE1}) has to be supplemented by equations governing the carrier dynamics.
In all models so far published this is done on the level of a diffusion equation for the excess carrier density $N$, 
\begin{equation}
\frac{\partial N}{\partial t}-\nabla\big[ D\nabla N\big]+R(N)+R_{\text{stim}}=\frac{j}{ed}
\end{equation}
assuming charge neutrality in the active region.
The thickness of the active region is denoted by $d$ and the rates of spontaneous (including non-radiative and radiative) and stimulated recombination by $R$ and $R_{\text{stim}}$, respectively, and $\nabla=(\partial/\partial x,\partial/\partial z)$.

The current density is given by \cite{joyce1980}
\begin{equation}
\label{currentdensity}
j(x,z)=
\begin{cases}
\big[U-U_{\text{F}}(x,z)\big]/r&\quad\text{if}\quad x,z\in \text{active stripe}\\
\big[\nabla^2U_{\text{F}}\big]/\Omega &\quad\text{elsewhere}
\end{cases}
\end{equation}
with $U$ being the bias and $r$ and $\Omega$ being the resistivity and the sheet resistance, respectively, of the p-doped layers.
The dependence of the current density within the active stripe on the Fermi voltage $U_{\text{F}}$, \textit{i.e.} the spacing of the electro-chemical potentials of electrons and holes,
results in a preferred current injection into regions with a low carrier density, 
thus counteracting spatial holeburning \cite{wuensche, eliseev1997}.
The reason is that $U_{\text{F}}$ decreases and thus the difference $U-U_{\text{F}}$ increases with decreasing $N$.

As shown in \cite{joyce1982}, the diffusion coefficient reads
\begin{equation}
D=\mu_p(p_0+N)\frac{dU_{\text{F}}}{dN}
\end{equation}
with $\mu_p$ being the mobility of the holes in the active layer and $p_0$ the equilibrium density of the holes.
For Boltzmann statistics, it is given by $D=2k_\text{B}T\mu_p/e$ ($k_\text{B}$ Boltzmann constant, $T$ temperature, $e$ elementary charge).

The rate of stimulated recombination is given by
\begin{equation}
\label{Rstim1}
R_{\text{stim}} = v_{\text{g,eff}}\,g_{\text{eff}} \left[|E^+|^2+|E^-|^2\right]
\end{equation}
with the effective gain 
\begin{equation}
g_{\text{eff}}=\frac{1}{\text{Re}(n_{\text{eff}})}\int n_{\text{r}}g\Phi^2dy.
\end{equation}
The output power $P_{0,L}$ at $z=0$ and $z=L$ is then given by
\begin{equation}
P_{0,L} = \hbar\omega_0 v_{\text{g,eff}} \,d (1-r_{0,L}^2) \int |E^{\mp}(x,z)|_{z=0,L}^2 dx.
\end{equation}
Note, that the dispersion of the gain must be included into (\ref{Rstim1}), too.

\subsection{Worked example}

\begin{figure}[!t]
\centering
\includegraphics[width=2.5in]{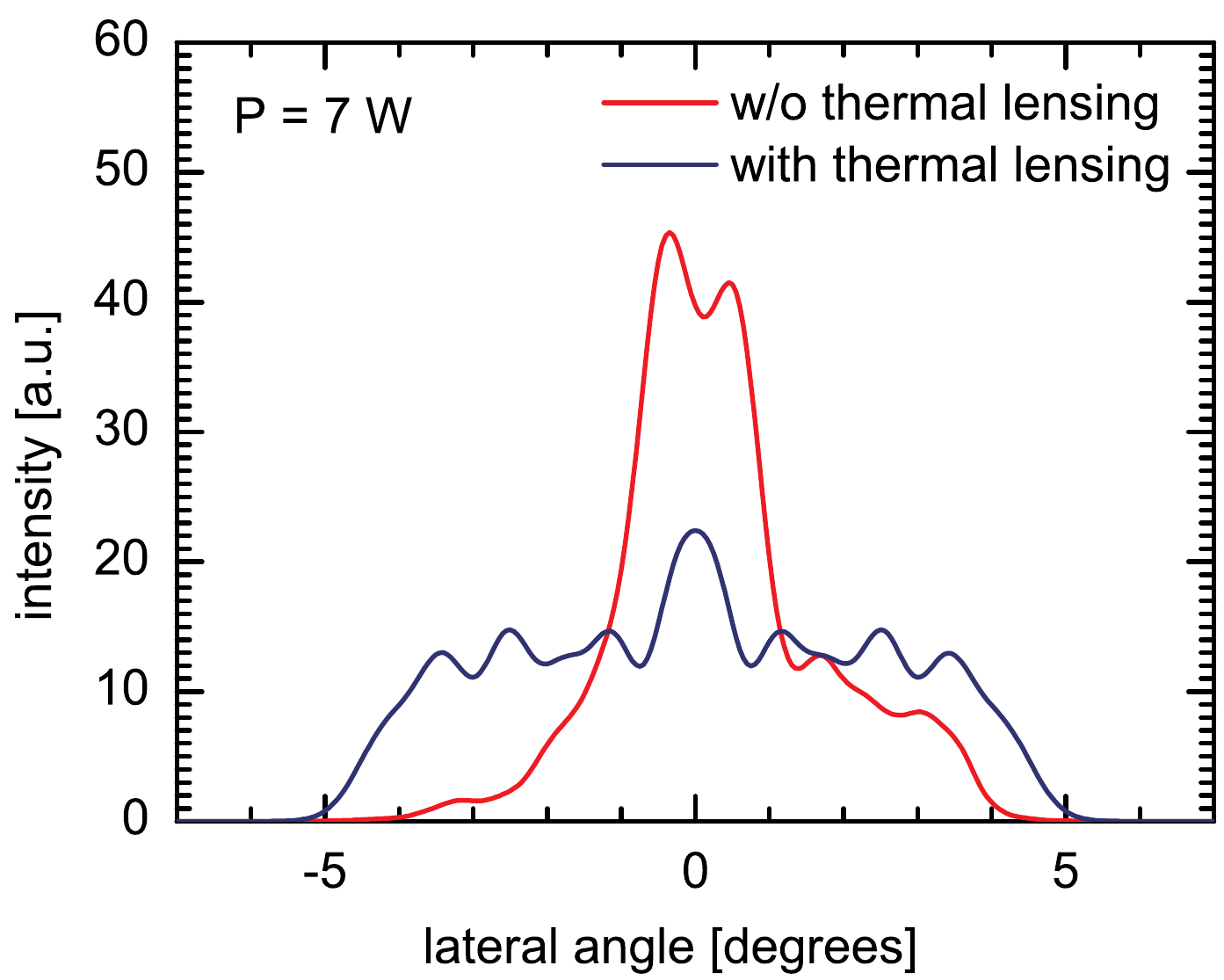}
\caption{Time-averaged lateral profiles of the far field intensity of a gain-guided BA DFB laser without and with the thermal lensing effect at an output power of $P=7~$W.}
\label{FFaverage}
\end{figure}

The multi-peaked and not diffraction-limited lateral field profile of 
wide-aperture semiconductor lasers has been a long standing problem 
and has been investigated in the past by numerous authors
\cite{thompson,mehuys,lang1991,marciante1996,blaaberg}.
Although the broadening of the far field of continuous-wave operating lasers with increasing power can be at least partially attributed to the thermal lensing effect \cite{moloney2000SPIE, wenzel2011, crump2012}, a complete picture of the origin and mechanism has yet to be revealed.

One mechanism is believed to be due to carrier induced antiguiding, \textit{i.e.} the reduction of
the refractive index with increasing carrier density.
This leads to a self-focusing mechanism because the index increases in regions of high intensity due to a depletion of the injected carrier density and can result in the formation of lasing filaments \cite{thompson}.

As stated in \cite{mehuys}, the maximum local increase of the real part of the effective index is given by $\text{Re}(\delta n_{\text{eff}})=\alpha_{\text{H}}g_{\text{eff,th}}/2k_0$ because the carrier density can not be depleted below transparency.
Hence, filamentation effects in state-of-the-art high-power strained quantum-well lasers having a low alpha factor $\alpha_{\text{H}}$ around $2$ and a small threshold gain $g_{\text{eff,th}}$ due low internal and outcoupling losses (large cavity length) should be of less importance compared to lasers investigated earlier.

Another mechanism that could explain the multi-peaked structure and the broadening of the farfield
is the simultaneous lasing of a large number of waveguide modes, originating from a built-in or thermally induced waveguide.
Indeed, recent experiments reveal, that even at currents several times above threshold the lateral modes can be clearly identified by spectrally-resolved near- and farfield measurements \cite{stelmakh2006, crump2012}.

\begin{figure}[!t]
\centering
\includegraphics[width=3in]{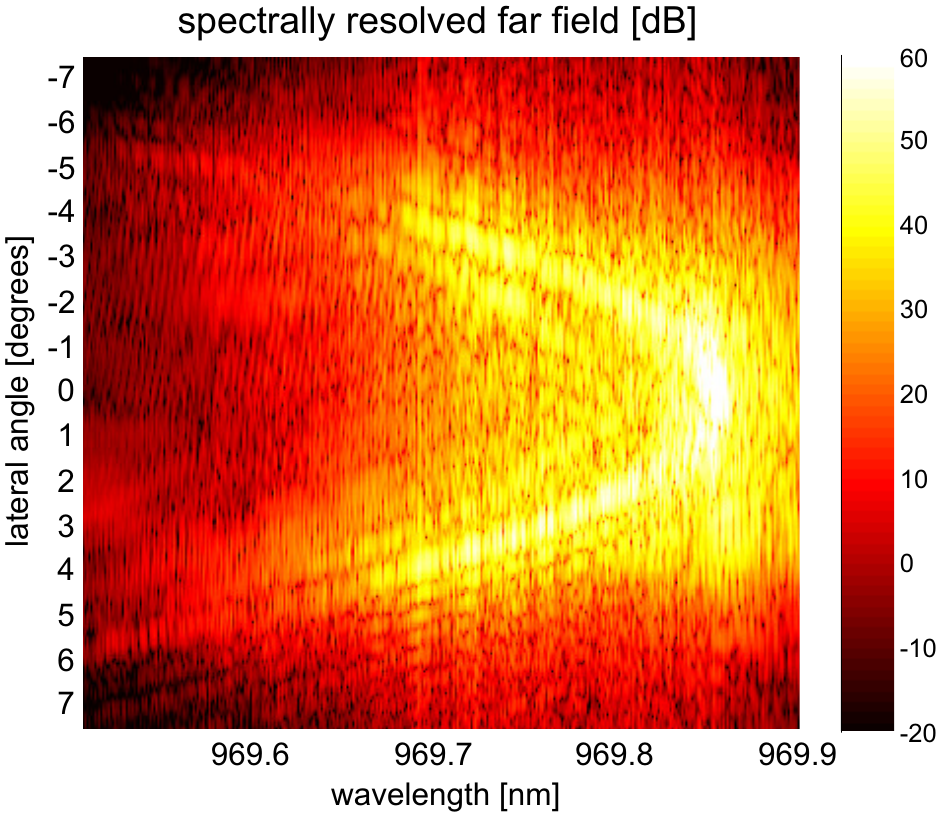}
\caption{Spectrally resolved far field profiles of the gain-guided BA DFB laser without the thermal lensing effect at an output power of $P=7~$W.}
\label{FFspectral}
\end{figure}

\begin{figure}[!t]
\centering
\includegraphics[width=3in]{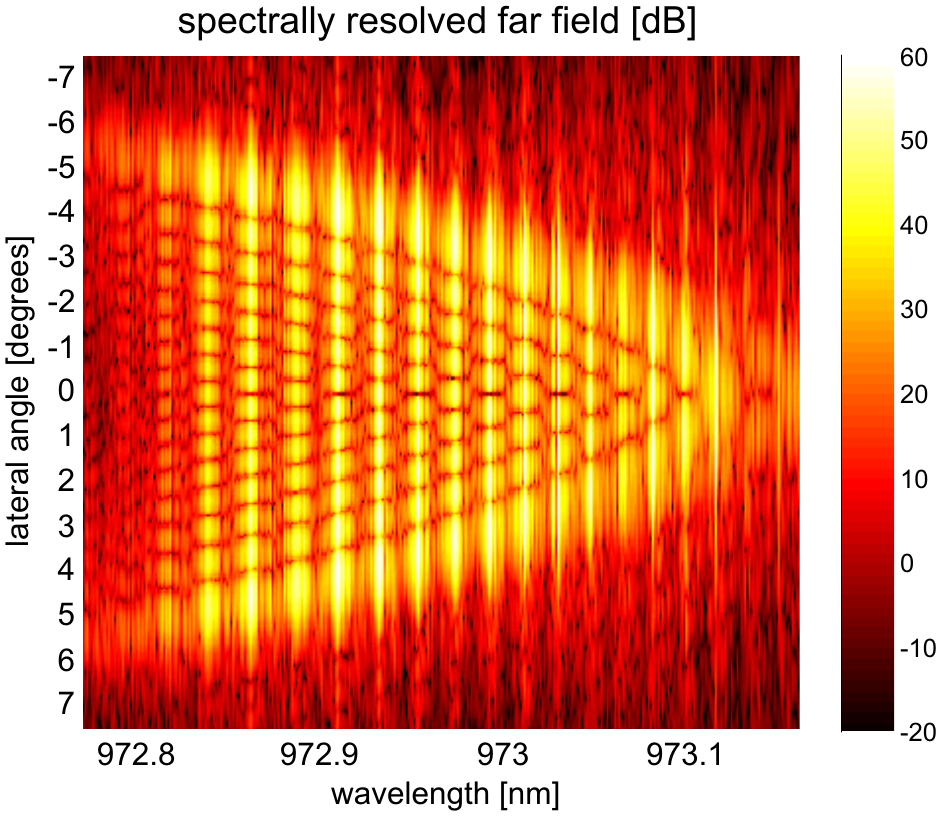}
\caption{Spectrally resolved far field profiles of the gain-guided BA DFB laser with the thermal lensing effect at an output power of $P=7~$W.}
\label{FFspectral2}
\end{figure}

In what follows we present exemplary results obtained with the simulation tool \lq\lq WIAS-LASER\rq\rq\ \cite{lichtner, spreemann2009, fiebig2010} for a broad-area distributed feedback (DFB) laser \cite{schultz2010}.
The device has a cavity length of $L=3~$mm, a total width of $500~\mu$m with an active stripe width of $W=90~\mu$m and a coupling coefficient of $\kappa=3.2~\text{cm}^{-1}$.

The discretization steps are $\Delta x=0.5~\mu$m and $\Delta z=5~\mu$m into the lateral and longitudinal directions, respectively.
The latter corresponds to a time step of $\Delta t=31.7~$fs. 
The simulation is performed over a total time window of $20~$ns, but temporal averaging and Fourier transformation is done during the last $12$~ns thus excluding the turn-on behavior.

The dependence of the effective gain and index change on the carrier density is modeled as
$g_{\text{eff}}(N)=g'\text{ln}\left(N/N_{\text{tr}}\right)$
and $\Delta n_{\text{eff}}(N)=-\sqrt{n'N}$,
respectively, with $g'=33~\text{cm}^{-1}, N_{\text{tr}}=1.4\cdot 10^{18}~\text{cm}^{-3}$,  $n'=2.5\cdot10^{-25}~\text{cm}^3$.
The square root like dependence of the index on the carrier density is motivated by 
the result of
microscopic calculations \cite{wenzel1999} 
and was firstly used in
\cite{borruel2004}.

We will consider a device without a built-in index step emitting a temporal averaged output power of about $P=7$~W at a current of $I=8~$A.
Fig. \ref{FFaverage} shows the temporal averaged profiles of the intensity of the far field
\begin{equation}
E_{FF}(\Theta,t)\propto \int E^-(x,z=0,t)\text{e}^{ik_0\text{sin}(\Theta)x}dx
\end{equation}
for two cases, without and with a temperature induced index profile due to self-heating. 
The stationary temperature profile has been calculated in advance by the tool JCMsuite \cite{wenzel2011, pomplun2012}.
If the thermal lens is not taken into account, the far field profile is asymmetric.
This could be caused by a symmetry-breaking mechanism inherent to purely gain-guided devices \cite{blaaberg}.
With the thermal-lensing effect the far field has a symmetric shape, but is more divergent.

A deeper insight can be gained from the spectrally resolved far fields
as shown in Figs. \ref{FFspectral} and \ref{FFspectral2}.
In Ref. \cite{moloney2000AIP} the parabolic shaped dispersion curve associated with a longitudinal mode, visible in Fig. \ref{FFspectral}, was misinterpreted as a sign of lateral filamentation.
We think that the opposite is true: The parabola indicates the surviving lateral mode structure, and its blurred  appearance reflects the filamentation.
Due to the lack of any carrier-density independent index profile, the modes react very sensitively to any changes of the carrier density, caused by local intensity fluctuations.

The index profile resulting from the self-heating stabilizes the lateral modes, so that they 
can be clearly identified in the spectrally resolved far field shown in Fig. \ref{FFspectral2} according to their number of nodes and antinodes.
The parabolic shaped dispersion curve is caused by the fact, that according to 
(\ref{ModenAbstand}) the lasing wavelengths of the modes decrease and that the far field divergence increases with rising mode index.
A similar interpretation was also given in \cite{boehringer2007}.
We should note, that there is only one longitudinal mode lasing due to the DFB operation. 

\begin{figure}[!t]
\centering
\includegraphics[width=3in]{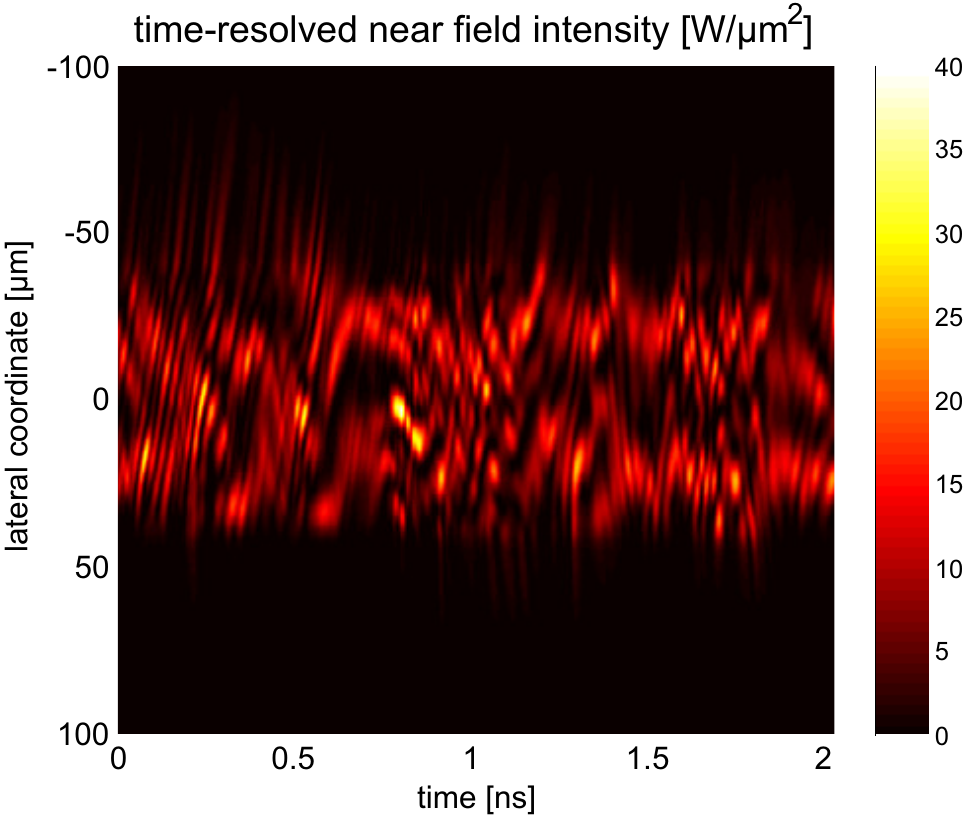}
\caption{Time-resolved near field profiles of the gain-guided BA DFB laser without the thermal lensing effect at an output power of $P=7~$W.}
\label{NFhistory}
\end{figure}

\begin{figure}[!t]
\centering
\includegraphics[width=3in]{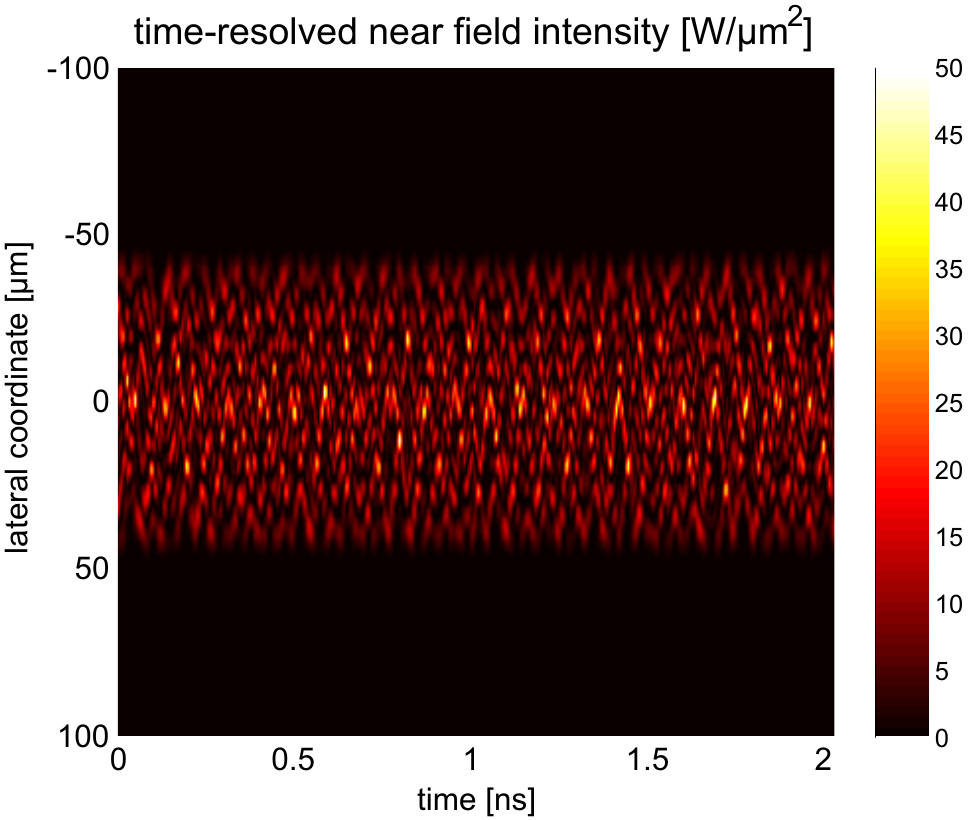}
\caption{Time-resolved near field profiles of the gain-guided BA DFB laser with the thermal lensing effect at an output power of $P=7~$W.}
\label{NFhistory2}
\end{figure}

The time-resolved near fields shown in Figs. \ref{NFhistory} and \ref{NFhistory2} reveal the non-stationary behavior, which was already found earlier \cite{fischer}.
It can be interpreted to be caused by mode beating in time and space. 
Obviously, in the purely gain-guided case (Fig. \ref{NFhistory}) less modes are involved compared to the case with a superimposed index profile (Fig. \ref{NFhistory2}).
The appearance of the bright spots can be thought to be the result of a constructive interference of the lateral modes.
Although the results are encouraging, further analytical and numerical investigations are needed in order to reveal the relative contributions of lateral mode and filamentary structures to the optical field of wide-aperture lasers and to obtain quantitative agreement with experimental results. 

\section{Stationary drift-diffusion based simulation}

\subsection{Basic equations}

The standard model \cite{wachutka1990,li1992,witzigmann2000} for the numerical evaluation of the distributions of the electron and hole densities $N$ and $P$, temperature $T$ and electron and hole densities $\vec{j}_n$ and $\vec{j}_p$ in semiconductor lasers is the energy transport model, which consist of the Poisson equation for the electro-static potential $\varphi$, 
\begin{equation}
\label{Poisson}
-\nabla\big[\varepsilon_{\text{s}}\nabla\varphi\big]=C-P+N
\end{equation}
with $\varepsilon_{\text{s}}$ being the static dielectric constant and $C$ the charged impurity density,
the continuity equations 
\begin{align}
\label{continuity}
\nabla\boldsymbol{j}_n &  =R+R_{\text{stim}}\\
-\nabla\boldsymbol{j}_p & =R+R_{\text{stim}}
\end{align}
and the heat flow equation 
\begin{equation}
\label{heat}
-\nabla\big[\kappa_{\text{L}}\nabla T\big]=H
\end{equation}
with $\kappa_{\text{L}}$ being the thermal conductivity.
The various implementations of the model differ in the assumptions for the current densities the $\vec{j}_n$, $\vec{j}_p$ and the heat source $H$.
Expressions in agreement with the thermodynamic principles can be found in Ref. \cite{bandelow}.
The rate of stimulated recombination is given here by
\begin{equation}
\label{rstim}
R_{\text{stim}}=\frac{n_{\text{r}}g}{\hbar\omega_0}\sum_\nu\frac{P_\nu|\chi_\nu|^2}{n_{\text{mod},\nu}\int|\chi_\nu|^2dxdy}
\end{equation}
with $\chi_\nu$ obtained by solving (\ref{waveguide}).

The equations are supplemented by proper boundary conditions \cite{bandelow}.
For example, the boundary condition for the heat flow equation reads
\begin{equation}
{\vec \nu}\kappa_{\text{L}}\nabla T=h\big[T_{\text{s}}-T\big]
\end{equation}
where ${\vec \nu}$ is the normal unit vector and $T_{\text{s}}$ is the heat sink temperature.
The heat transfer coefficient $h$ models the heat flow to the exterior region.

The optical output power for a given applied bias $U$ is obtained by a solution of (\ref{TME1}).
However, in Fabry-Perot lasers it is sufficient to determine the forward and backward propagating optical power $P_\nu^{+}=|A_\nu^+|^2$ and $P_\nu^{-}=|A_\nu^-|^2$, which are solutions of the equations
\begin{equation}
\label{power1}
\pm\frac{dP^\pm_\nu}{dz}=\left[2\text{Im}\Big(\beta_\nu+\frac{\Omega_\nu}{v_{\text{g},\nu}}\Big)-\alpha_0\right]P^\pm_\nu
\end{equation}
with $P_\nu=P_{\nu}^++P_{\nu}^-$.
The loss coefficient $\alpha_0$ includes all loss mechanisms which are not already accounted for in the imaginary part of $\beta$, such as additional internal or scattering losses.
The equations (\ref{power1}) have to be solved subject to the boundary conditions
\begin{equation}
\label{boundary}
P^+_\nu(0)=R_0P^-_\nu(0)\quad\text{and}\quad P^-_\nu(L)=R_LP^+_\nu(L).
\end{equation}
The lasing wavelength can be approximately determined by determining the maximum of the integral
$\int_0^L\text{Im}\big(\beta_\nu+\Omega_\nu/v_{\text{g},\nu}\big)dz$.

Eqs. (\ref{TME1}) or (\ref{power1}) can be conveniently solved by the \lq\lq Treat Power as a Parameter\rq\rq\ (TPP) method as introduced in \cite{wuensche} and used in \cite{wenzel1996SPIE}.
It is based on the observation, that the relative propagation factor $\Delta\beta_{\nu}$ in (\ref{TME1}) or imaginary part  $\text{Im}(\beta_\nu)$ in (\ref{power1}) can be considered to be a function of the bias, the local power and the unknown wavelengths of the lasing and nonlasing modes.
Thus, in a first step one calculates $\Delta\beta_{\nu}$ as a function of $U$, the power of the lasing waveguide mode $P_{\nu_l}=P_{\nu_l}^++P_{\nu_l}^-$, and the wavelengths by solving the drift-diffusion and waveguide equations in the transverse cross section and stores the results in a in a look-up table.
In a second step, (\ref{power1}) is solved by interpolating in the look-up table.
For the lasing waveguide mode, $\text{Im}(\Omega_{\nu_l})$ has to vanish and one has to determine the output power and the wavelength for given $U$.
For the non-lasing waveguide modes one has to determine $\Omega_\nu$ for the given $U$ and the power distribution of the lasing mode.

Longitudinal spatial hole burning (LSH) is included automatically via the power dependence of $\text{Im}(\beta_\nu)$ in Eq. (\ref{power1}).
If $\text{Im}(\beta_\nu)$ is evaluated at the average power 
$\bar{P}_{\nu_l}=\int\big[P_{\nu_l}^++P_{\nu_l}^-\big]dz/L$
in the cavity, the usual model neglecting LSH is recovered, because (\ref{power1}) is linear and can be 
analytically solved.

\begin{figure}[!t]
\centering
\includegraphics[width=2.8in]{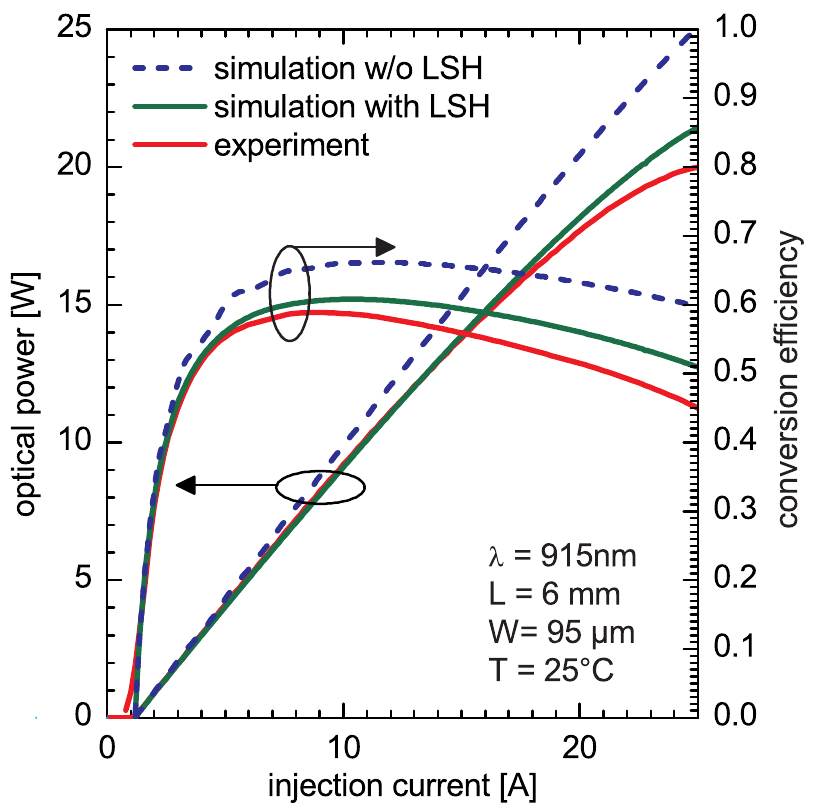}
\caption{Measured (solid red) and simulated optical output power (left axis) and conversion efficiency (right axis) versus injection current. The simulations were done without (dashed blue) and with (solid green) longitudinal spatial holeburning.}
\label{P_etac_I_sim_exp}
\end{figure}

\subsection{Worked example}
\label{4a}

In what follows we present results obtained with the simulation tool \lq\lq WIAS-TeSCA\rq\rq\ \cite{tesca} for a broad-area laser.
The device with a cavity length of $L=6~$mm and an active stripe width of $W=95~\mu$m has a double quantum well (DQW) InGaAs/GaAsP active region embedded into $2.2~\mu$m wide Al$_{0.30}$Ga$_{0.70}$As confinement and Al$_{0.85}$Ga$_{0.15}$As cladding layers.
The optical confinement factor of the DQW totals 1.8\%.
We performed a one-dimensional simulation in the transverse cross section neglecting any lateral effects as outlined
in \cite{wenzel2009oqe} and \cite{wenzel2010njp}. The pre-factor for the gain and the Shockley-Read-Hall recombination life times (
$\tau_n=\tau_p=1.3~\text{ns}$
identical for all layers) have been fitted on the results of length-dependent pulsed measurements of threshold current and slope efficiency of uncoated devices. 
The cross-sections of free-carrier absorption have been chosen to $f_{\text{cn}}=3\times 10^{-18}~\text{cm}^{2}$ and $f_{\text{cp}}=7\times 10^{-18}~\text{cm}^{2}$ for electrons and holes, respectively.

Fig. \ref{P_etac_I_sim_exp} shows the measured and simulated power-current characteristics of the laser.
Experimentally, a maximum output power of $P = 20~$W at a current of $I = 25~$A was achieved limited by thermal rollover. 
The experimental data were reproduced theoretically in a two-stage process with and without LSH. 
The reason for the rollover of the characteristics occurring already without LSH has been explained in detail in \cite{wenzel2010njp}.
With increasing current and power, the reduction of the gain caused by the
temperature rise 
($\Delta T = 42~$K at $I = 25~$A)
must be compensated for by a corresponding increase in the carrier densities
in the active region, which leads as a consequence also to an increase in the carrier densities in the confinement layers. 
Another effect is the bending of the quasi-Fermi energy of the holes with increasing applied bias and, as a consequence, the bending of the band-edge energies due to the voltage drop in the p-doped layers. 
This leads to a corresponding linear increase in the electron density with increasing distance from the active region
up to $n\approx 3\times10^{16}~\text{cm}^{-3}$.
The increased carrier densities in the active region
and in the bulk layers give rise to enhanced non-stimulated recombination and free-carrier
absorption \cite{ryvkin2005a,ryvkin2005b}.

However neglecting LSH, not only the roll-over power, but also the conversion efficiency is exaggerated which is due to the overestimation of the slope-efficiency. 
Good agreement between measurement and simulation is only achieved if LSH is included,
as Fig. \ref{P_etac_I_sim_exp} reveals.

\begin{figure}[!t]
\centering
\includegraphics[width=2.8in]{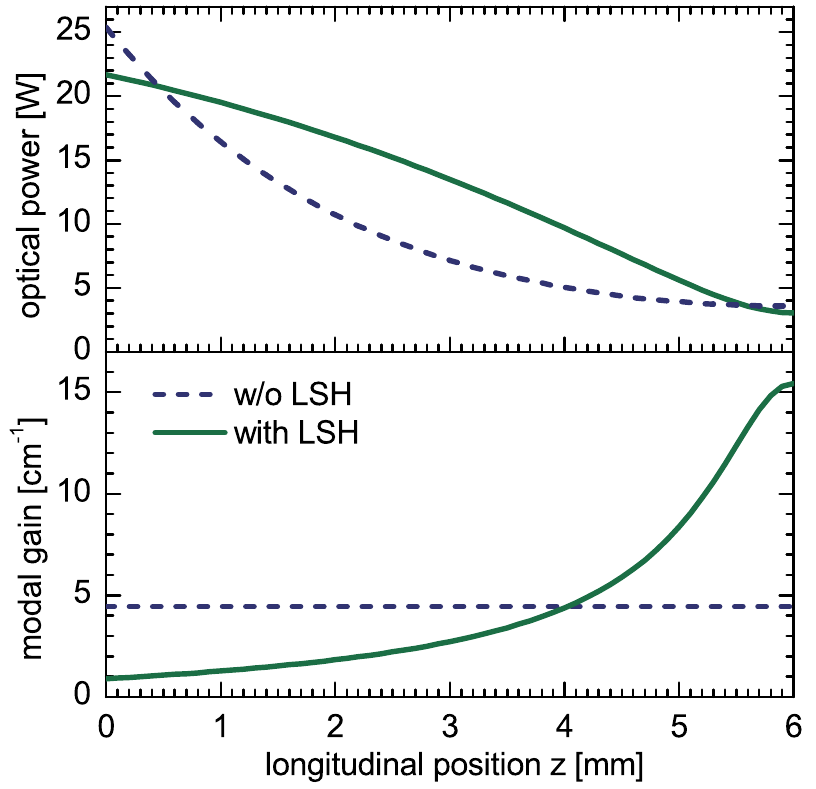}
\caption{Longitudinal profiles of optical power (top) 
and modal gain 
(bottom) at an injection current of $I=25~$A without (dashed blue) and with (solid green) longitudinal spatial holeburning.}
\label{P_g_z}
\end{figure}

To illustrate the mechanism of LSH, Fig. \ref{P_g_z} shows the longitudinal profiles of the optical power, the modal
gain and the injected current density for the two models at a current of $I=25~$A.
The rear facet with a reflectivity of $R_L=0.98$ is located at z$ = 6$~mm and the front facet 
with a reflectivity of $R_0=0.005$ at $z = 0$.
LSH leads to a strong depletion of the gain in the vicinity of the front facet, 
which is compensated by a corresponding rise of the gain at the rear facet.
This results in a weaker increase of the power towards the front facet and hence a lower power at $z=0$, \textit{i.e.} a lower output power, and is in contrast to Ref.
\cite{ryvkin2011} 
where based on an analytical model it was found that the values of the power at the facets don't dependent 
on the fact, whether LSH is considered or not.

It should be noted, that even with LSH there is still a small deviation in roll-over power and conversion efficiency between measurement and simulation.
One reason is the neglect of the series resistances of the p-contact and the substrate in the simulation, which lead to an
additional voltage drop reducing the conversion efficiency but also to additional heating reducing the roll-over power.

\subsection{Heterojunctions}

Heterojunctions characterized by discontinuities in the edges of the conduction and valance bands as well as abrupt changes of effective masses and mobilities require special attention.
If the validity of the thermodynamic based drift-diffusion approach is assumed, the electro-chemical potentials  
should be continuous through the heterojunctions.
The reason is, that the electro-chemical potentials and the inverse temperature
can be interpreted as Lagrange multipliers in the functional for the maximization of the entropy subject to the constraints of charge and energy conservation \cite{albinus2002}.

The widely used thermionic emission theory which establishes additional conditions for the current densities \cite{li1990,grupen1991} is restricted to situations where the current essentially flows perpendicular to the junction and cannot be applied to intersections of three or more materials \cite{steiger}.
In high-power lasers there is no need to apply the thermionic emission theory with its difficulties because 
typically all heterojunctions (except at the QWs) are graded in order to avoid any drops of the electro-static potential.

\subsection{Quantum effects}

In the simulations presented in subsection \ref{4a}, the carriers in all layers were treated within the
drift-diffusion approach, neglecting quantum effects.
For more accurate simulations electrons and holes confined in active QWs must be treated in a special manner.
As outlined in detail in \cite{grupen, witzigmann2000, borruel2003, liu2005, steiger}, the transport of the confined carriers in the in-plane directions $(x,z)$ can be described by classical drift-diffusion, but in the perpendicular direction $y$ the carriers have to be described by their quantum-mechanical wave functions. 
Therefore, the carriers population partitions into carriers which have sufficient energy to be be considered as
unconfined and those which are confined.
The scattering between both populations is described by a capture rate which has to be included into the continuity equations (\ref{continuity}) as a recombination rate for the unconfined and a generation rate for the confined carriers.

It should be noted that the statistics of both carrier populations is governed by different electro-chemical potentials.
Whereas the quasi-chemical potentials of the unconfined carriers depend directly on the bias applied to the contacts, 
the quasi-chemical potentials of the unconfined carriers are determined by the capture rates.
Coupling of the populations takes also place via the Poisson equation (\ref{Poisson}).

Although the sketched picture where carriers are described quantum-mechanically in one direction and classically in the others has been included in several simulation tools, there are still open questions which have to be solved.
The energy which partitions the carrier populations can not be always clearly defined, for example in the
case of a QW located in an unbiased pn-junction.  
Another issue is the correct treatment of the transport in multi-quantum well structures between the QWs through the barriers.

The impact of the non-equilibrium between confined and unconfined carriers in high-power lasers on the internal efficiency, \textit{e.g.}, has been investigated until now using only rate-equation approaches \cite{tansu2005}, not the drift-diffusion model.

\section{Summary and Outlook}
We presented models for the calculation of passive cavity modes, for the investigation of the spatiotemporal behavior of the optical field and of the stationary simulation of the light-current characteristics of high-power semiconductor lasers.
Future work should be directed at full three-dimensional calculation of the optical field and an improvement of the physical models underlying the time-dependent active cavity simulations, including a better description of the carrier and heat transport by the energy transport model instead of the diffusion equation and stationary heat conduction equation.
Finally, we would like to mention that the dependence of many material parameters and functions such as mobilities and absorption coefficients on composition, temperature and doping is not well known and needs to be improved.

\appendix[Proof of mode orthogonality]

Let us write down (\ref{DTWE}) for 2 modes with indices $m$ and ${m'}$:
\begin{equation}
\label{mplus}
\left[\frac{k_0nn_{\text{g}}\Omega_m}{c\beta_0}
- i\frac{\partial }{\partial z}
+ \frac{1}{2 \beta_0}\mathbf{\Delta_t} 
+ \Delta\beta \right]E_m^{+} 
+ \frac{k_0^2\xi^{+}}{2 \beta_0}E_m^{-}
= 0
\end{equation}
\begin{equation}
\label{mminus}
\left[\frac{k_0nn_{\text{g}}\Omega_m}{c\beta_0}
+ i\frac{\partial}{\partial z}
+ \frac{1}{2 \beta_0}\mathbf{\Delta_t}
+ \Delta\beta \right]E_m^{-} 
+ \frac{k_0^2\xi^{-}}{2 \beta_0}E_m^{+}
= 0
\end{equation}
\begin{equation}
\label{nplus}
\left[\frac{k_0nn_{\text{g}}\Omega_{m'}}{c\beta_0}
- i\frac{\partial}{\partial z}
+ \frac{1}{2 \beta_0}\mathbf{\Delta_t} 
+ \Delta\beta \right]E_{m'}^{+} 
+ \frac{k_0^2\xi^{+}}{2 \beta_0}E_{m'}^{-}
= 0
\end{equation}
\begin{equation}
\label{nminus}
\left[\frac{k_0nn_{\text{g}}\Omega_{m'}}{c\beta_0}
+ i\frac{\partial }{\partial z}
+ \frac{1}{2 \beta_0}\mathbf{\Delta_t} 
+ \Delta\beta \right]E_{m'}^{-} 
+ \frac{k_0^2\xi^{-}}{2 \beta_0}E_{m'}^{+}
= 0
\end{equation}
Next 
(\ref{mplus}) is multiplied with $E_{m'}^-$, 
(\ref{mminus}) with $E_{m'}^+$, 
(\ref{nplus}) with $-E_m^-$,
and (\ref{nminus}) with $-E_m^+$.
Then all equations are integrated over the cavity and added to obtain
\begin{multline}
\label{sum}
\frac{k_0\left(\Omega_m-\Omega_{m'}\right)}{c\beta_0}\int nn_{\text{g}}\left[E_m^{+}E_{m'}^{-}+E_m^{-}E_{m'}^{+}\right]dxdydz\\
-i\int\frac{\partial}{\partial z}\left[
E_m^{+}E_{m'}^{-}-E_m^{-}E_{m'}^{+}
\right]dxdydz\\
+\frac{1}{2 \beta_0}\int\nabla_t\big[
E_{m'}^{-}\nabla_t E_m^{+}+E_{m'}^{+}\nabla_t E_m^{-}\\
-E_m^{-}\nabla_t E_{m'}^{+}-E_m^{+}\nabla_t E_{m'}^{-}
\big]dxdydz=0
\end{multline}
where the derivatives have been factored out.
The second and third terms vanishes due to the boundary conditions (\ref{refl})-(\ref{TBC}).
Hence, the first term has to vanish, too, with leads to (\ref{Orthogonality}).

\section*{Acknowledgment}
The author would like to thank R. G\"uther  and P. Crump (FBH) for a critical reading of the manuscript and M. Radziunas and U. Bandelow (WIAS) for providing the simulation tool \lq\lq WIAS-LASER\rq\rq.

\ifCLASSOPTIONcaptionsoff
  \newpage
\fi



\begin{IEEEbiographynophoto}{Hans Wenzel}
received the Diploma and Doctoral degrees in physics from Humboldt-University Berlin, Germany, in 1986 and 1991, respectively.
His thesis dealt with the electro-optical modeling of semiconductor lasers.  
From 1991 to 1994, he was involved in a research project on the simulation of distributed feedback lasers. 
In 1994, he joined the Ferdinand-Braun-Institut, Leibniz-Institut f\"ur H\"ochstfrequenztechnik, where he is engaged in the development of high-brightness semiconductor lasers. 
He authored or co-authored more than 250 journal papers and conference contributions.
His main research interests include the analysis, modeling, and simulation of optoelectronic devices.
\end{IEEEbiographynophoto}

\end{document}